\documentclass[
 aps,
 reprint,
 amsmath,amssymb,
]{revtex4-2}

\usepackage{graphicx}% Include figure files
\usepackage{dcolumn}% Align table columns on decimal point
\usepackage{bm}% bold math
\usepackage{dsfont}
\usepackage{amsthm}
\usepackage{amsmath}
\usepackage{amsfonts}
\usepackage{tensor}
\usepackage{mathtools}
\usepackage{amssymb}
\usepackage{bm}
\usepackage{setspace}
\usepackage{enumitem}
\usepackage{mdwlist}
\usepackage{parskip}
\usepackage{listings}
\usepackage{color}
\usepackage{hyperref}
\usepackage{subcaption}
\usepackage[normalem]{ulem}

\usepackage[mathlines]{lineno}% Enable numbering of text and display math
\newcommand{\dif}{\mathrm{d}}
\newcommand{\avg}[1]{\left\langle #1 \right\rangle}

\DeclarePairedDelimiter\abs{\lvert}{\rvert}%
\begin{document}

\preprint{APS/123-QED}

\title{Coulomb drag between two strange metals}% Force line breaks with \\

\author{E. Mauri}
 \email{E.Mauri@uu.nl}
\author{H.T.C. Stoof}%
 \email{H.T.C.Stoof@uu.nl}
\affiliation{%
Institute for Theoretical Physics and Center for Extreme Matter and Emergent Phenomena, Utrecht University, Princentonplein 5, 3584 CC Utrecht, The Netherlands}


\begin{abstract}
We study the Coulomb drag between two strange-metal layers using the Einstein-Maxwell-Dilaton model from holography. We show that the low-temperature dependence of the drag resistivity is $\rho_D \propto T^4$, which strongly deviates from the quadratic dependence of Fermi liquids. We also present numerical results at room temperature, using typical parameters of the cuprates, to provide an estimate of the magnitude of this effect for future experiments. We find that the drag resistivity is enhanced by the plasmons characteristic of the two-layer system. %(\textcolor{red}{620 characters. Official limit 600})
%\begin{description}
%\end{description}
\end{abstract}

\maketitle

\section{Introduction}\label{sec:introduction}
The cuprates have been of interest to both the experimental and theoretical physics community for the past three decades due to their peculiar metallic properties that cannot be explained within the standard Fermi-liquid framework \cite{Proust2019, Keimer2015}.
In this class of materials, that have in common a layered structure of $\text{CuO}_2$ planes, strong Coulomb interactions between the electrons give rise to unusual and still unclear quantum behavior, of which the high-temperature superconducting phase is of most technological importance. 
Above the maximum critical temperature of the superconducting phase, we find the so-called strange metal, whose anomalous linear-in-$T$ resistivity \cite{Custers2003, Cooper603, bruin_similarity_2013, Analytis2014, Legros2018, Licciardello2019} characterizes its non-Fermi-liquid behavior, even if the superconductivity is suppressed by a magnetic field \cite{Hussey_2018}. Given the importance of understanding the physics at play in the strange-metal phase in order to also get a better grasp on the instability towards high-temperature superconductivity, there have been a variety of attempts to model the strongly interacting cuprates \cite{Lee2006, tJ_review, AFM_multilayer, marginal_FL, marginal_FM_2, Zaanen1996, Emery1999, Berg_2009, Matthias_2009, Zhang68}. One technique, in particular, is the application of the AdS/CFT correspondence, or holographic duality \cite{Maldacena1998}, conjectured and first developed by the string-theory community, but that has proven to be a powerful tool to qualitatively study low-energy properties of strongly interacting quantum matter as well \cite{hartnoll2016holographic, Zaanen2015, ammon_erdmenger_2015}. In fact, the more easily tractable classical limit of a gravitational theory 
allows for an effective description of the generic behavior of a strongly interacting system in one lower spatial dimension, such as a cuprate strange metal layer with its strong (on-site Hubbard $U$) electromagnetic repulsion.

In this paper we use a holographic Einstein-Maxwell-Dilaton model (EMD) of the strange metal \cite{Gubser_2010} to study a two-layer system where the interaction among the different layers is only mediated by the long-range Coulomb force. The importance of such a setting is motivated by the experimental relevance of multi-layered systems. For simplicity most of the holographic studies of the cuprates focus on describing the behavior of a single-layer, however, there has been recent experimental interest in probing also the bulk response of the cuprates \cite{Hepting2018, PhysRevLett.125.257002}. Experimental results in both hole- and electron-doped cuprates show that the density response is governed by the inter-layer Coulomb interaction, giving rise to low-energy acoustic plasmon modes. Plasmons are ubiquitous in condensed-matter materials due to the electric charge of the electron, but only recently a way to capture dynamical Coulomb interactions and, hence, screening effects due to the charged nature of the system, has been proposed in holography \cite{Gran2017, Aronsson2018, Gran2019, Mauri2019, Anomalous_attenuation}, opening the door to the possibility of studying `holographic' plasmons and providing a simplified framework to test experimental results on acoustic plasmons in multi-layered cuprates \cite{Mauri2019}.

\begin{figure}[b]
\includegraphics[width=0.4\textwidth]{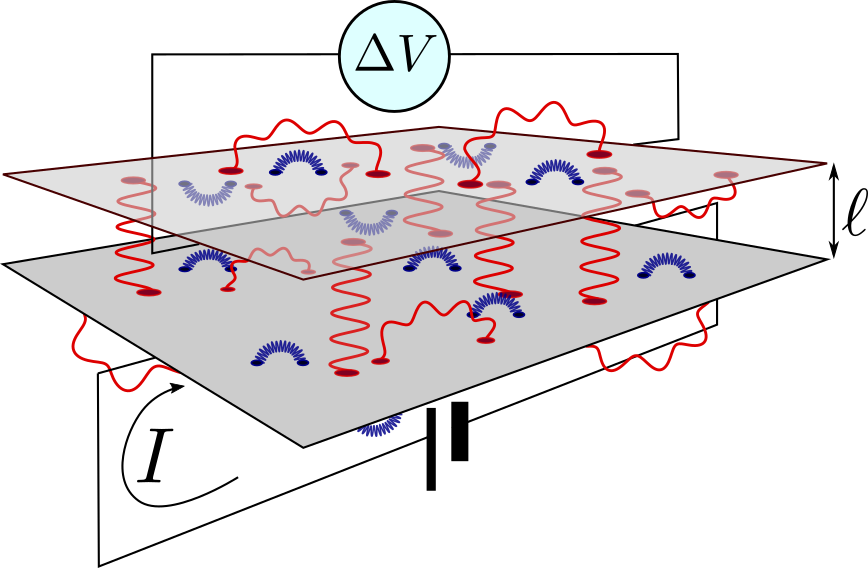}
\caption{\label{fig:set-up}
Experimental set-up for a Coulomb drag measurement. A current $I$ flows through the active layer and the voltage drop $\Delta V = - R_D I$ induced in the passive layer is observed to measure the drag resistance. In the holographic description under study we have two types of interactions, i.e., strong in-plane interactions leading to strange-metal behavior (blue) and the long-range Coulomb interactions (red). }
\end{figure}

In view of these developments, our main aim is here to present a link between holographic predictions and experimental results by studying the density response of two holographic strange-metal layers. In particular, by considering the Coulomb drag resistivity, we are able to obtain also  quantitative estimates of the drag effect in the hope to provide a verifiable prediction to either validate the holographic model or to show its shortcomings. Here Coulomb drag refers to a transport phenomenon of a system of closely spaced but electrically isolated layers, where a current driven through the `active' layer gives rise to a current in the nearby `passive' layer due to inter-layer scattering of the charge carriers mediated by the long-range Coulomb interaction that couples density fluctuations among the different layers \cite{RevModPhys.88.025003}. A typical experimental set-up is shown in Fig.~\ref{fig:set-up}, where a known current is driven through one layer and the drag effect, parametrized by the drag resistivity $\rho_D$, is studied by measuring the voltage drop in the passive layer.

In the holographic description we are using, the system is characterized by two kinds of interactions, a strong one that is of short range and therefore does not contribute to the inter-layer scattering, and the long-range Coulomb interaction that instead also gives rise to the coupling between the separate layers. The success of the holographic duality relies on the fact that it naturally describes a strongly interacting system, such as a strange metal \cite{Hartnoll2009}, and we thus can take two copies of such a model to describe the two layers. On the other hand, it is important to realize that holographic models present always a `neutral' response, where there are no photons in the system. However, by performing a so-called double-trace deformation \cite{Witten2001, Witten2003, Marolf_2006}, we have recently shown how to add within holography also the long-range Coulomb screening effects to multi-layered models on top of the strong neutral interactions \cite{Mauri2019}. We use this procedure here to couple the holographic strange-metal layers in order to study the Coulomb drag resistivity, as we explain in detail in the following.

\section{Einstein-Maxwell-Dilaton theory}\label{sec:emd_theory}
Let us briefly review how we obtain the strange-metal response function for a single layer from holography. The holographic duality states the equivalence between a strongly interacting quantum theory and a classical gravitational theory with one additional spatial dimension, where the coordinate $r$ of this additional dimension parametrizes the energy scale of the quantum field theory. In particular low-energy physics is described by the geometry in the deep interior ($r \to 0$) of the spacetime, while the near-boundary geometry ($r \to \infty$) characterizes the ultra-violet details of the dual field theory.

As a holographic model of a single-layer cuprate strange metal, we use a $(3+1)$-dimensional Einstein-Maxwell-Dilaton theory proposed by Gubser \textit{et al.} \cite{Gubser_2010}. This is one model in a class of holographic theories dual to a quantum field theory characterized at low energies by a `semi-local' quantum critical phase with a dynamical critical exponent $z = \infty$,  with the important implication that the scaling of the electron self-energy is then dominated by its frequency dependence $\hbar \Sigma(\omega, \bm k) \simeq  \omega(-\omega^2)^{\nu_k - 1/2}$, with momentum dependence only entering through the exponent $\nu_k$. The relevance of this result is due to the fact that, upon tuning the adjustable parameters in the holographic model, the holographic result reproduces the `power-law liquid' model, with $\hbar \Sigma''(\omega, \bm k) \propto  \omega^{2 \alpha}$ that is found to accurately describes the experimentally observed electron self-energy in angle-resolved photoemission spectroscopy (ARPES) measurements near the Fermi surface \cite{Reber_2015}. Here, $\alpha$ is a doping-dependent constant, with $\alpha = 1/2$ at optimal doping and the holographic model reducing to the famous marginal Fermi liquid \cite{marginal_FL, marginal_FM_2} proposed for the qualitative description of the optimally-doped strange metal. Moreover, the momentum-dependence in the exponent predicted by the Gubser-Rocha model used here has been recently shown to accurately describe deviations from the power-law liquid model away from the Fermi surface in high-quality ARPES measurements \cite{Smit2021, Mauri2022}, providing further support for the usefulness of the model in studying the response function of the strange metal. On the other hand, it is also important to keep in mind that the strange metal is further characterized by an anomalous scaling of some quantities at nonzero magnetic field, such as the Hall angle, and, while of no relevance for the results presented in this paper, the holographic model used here might fail to reproduce such a scaling \cite{Blake2015}. 

The gravitational action for the model is 
\begin{align}\label{eq:background_action}
\begin{split}
    S_{\text{EMD}} = \frac{c^3}{16\pi G} \int& \dif r \dif (c t) \dif^2 x \sqrt{-g}\left[R - \frac{(\partial_\mu\phi)^2}{2}\right. \\&+\left. \frac{6}{L^2} \cosh\left(\frac{\phi}{\sqrt{3}}\right) - \frac{e^{\phi/\sqrt{3}}}{4 g_F^2} F_{\mu\nu}^2 \right]\\
    &+ S_{\text{ct}}\text{ ,}
    \end{split}
\end{align}
with $r$ denoting the additional spacial direction of the curved bulk spacetime, $R$ the Ricci scalar, $\phi$ a dimensionless scalar field known as the dilaton \cite{Charmousis2010, Gouteraux2011}, and $F_{\mu\nu}$ the electromagnetic tensor with coupling constant
\begin{align}
    g_F^2 = \frac{c^4 \mu_0}{16 \pi G} \text{ ,}
\end{align}
where $[\mu_0] = \text{m kg}/\text{C}^2$ is a constant with the dimension of a magnetic permittivity. Finally, $L$ is the anti-de Sitter (AdS) radius and $S_{\text{ct}}$ contains the boundary counterterms necessary to have a well-defined boundary problem, and that are specified in the appendix. The dilaton field is such that the emergent low-energy theory is described by a semi-local quantum critical phase with both the dynamical critical exponent $z = \infty$ and the hyperscaling violation exponent $\theta = -\infty$ divergent but with a fixed ratio equal to $-1$. As previously mentioned, the dynamical exponent $z = \infty$ implies that at energies well below the Fermi energy, quantum critical correlations can only depend on momentum through the power of the frequency dependence, while $\theta/z = -1$ ensures that the entropy goes linearly to zero at zero temperature \cite{hartnoll2016holographic}. Ultimately, this scaling allows for the description of a quantum critical phase with a linear-in-$T$ resistivity and entropy \cite{Anantua2013} and an electron self-energy that is dominated by its frequency dependence as desired for a strange metal \cite{Gubser2012}. On the field-theory side, the dilaton is dual to a scalar operator $\mathcal{O}$ sourced by $\phi^{(0)}$, that leads to a nonzero trace of the energy-momentum tensor according to $\avg{T\indices{^\mu_\mu}} = \phi^{(0)}\avg{\mathcal{O}}$. This in turn implies, as we will see, that the speed of sound of the low-energy excitations of the system differs from that of a conformal invariant liquid.

For computational purposes it is convenient to re-write the action in Eq.~\eqref{eq:background_action} in dimensionless quantities by expressing everything in terms of physical constants and the dimensionful scale $L$ of the theory, that is by measuring distances in units of $L$ and energies in terms of $\hbar c/L$. We, therefore, define the dimensionless coordinates $(\tilde r, \tilde t, \tilde{\bm x}) \equiv (r, ct, \bm x)/L$, and we absorb the gauge coupling into the field $\tilde A_{\mu} \equiv A_\mu/g_F$. The action then becomes
\begin{align}\label{eq:dimless_background_action}
 \begin{split}
    \tilde S_{\text{EMD}} = \frac{c^3 L^2}{16\pi \hbar  G} \int& \dif \tilde r\, \dif \tilde t\, \dif^2 \tilde{x}\sqrt{-g} \left[R - \frac{(\partial_\mu\phi)^2}{2} \right. \\&+\left. 6 \cosh\left(\frac{\phi}{\sqrt{3}}\right) - \frac{e^{\phi/\sqrt{3}}}{4} \tilde{F}_{\mu\nu}^2 \right] \text{ .}
 \end{split}
\end{align}
From now on, we will mostly use dimensionless quantities and we will drop the tilde for notational convenience and explicitly state when we write expressions in terms of dimensionful quantities. Moreover, we follow the convention of setting the prefactor of the integral, $N_G \equiv c^3 L^2/16\pi \hbar G$, to unity for computational convenience, but we will comment extensively on its role in fixing the plasma frequency in a later section. This prefactor is related to the large-$N$ number of species of the boundary quantum field theory \cite{Zaanen2015}. 

The thermodynamics of this holographic strange-metal model is described by a solution to the coupled set of Einstein equations for the metric $g_{\mu\nu}$, Maxwell equations for the $U(1)$ gauge field $A_\mu$, and the Klein-Gordon equation for the real scalar field $\phi$ as obtained from the variation of the action:
\begin{align}
    &R_{\mu\nu} - \frac{1}{2} R g_{\mu\nu} = T_{\mu\nu}\text{ ,~~}
    \nabla_\mu(e^{\phi/\sqrt{3}} F^{\mu\nu}) = 0 \text{ ,} \nonumber\\
    &\nabla_\mu\nabla^\mu\phi = \frac{e^{\phi/\sqrt{3}}}{4\sqrt{3}}F^2 - \frac{2 \sqrt{3}}{L^2} \sinh(\phi/\sqrt{3}) \text{ ,}\label{eq:scalar}
\end{align}
where $R_{\mu\nu}$ and $R$ are the Ricci tensor and scalar, respectively, and the energy-momentum tensor %of the `matter' fields
equals
\begin{align}
    T_{\mu\nu}=& \frac{1}{2} \partial_\mu\phi \partial_\nu\phi + \frac{e^{\phi/\sqrt{3}}}{2} F_{\mu}^{~\sigma} F_{\nu \sigma}\\ \nonumber
    &- g_{\mu\nu} \left(\frac{e^{\phi/\sqrt{3}}}{8} F^2 + \frac{(\partial_\sigma \phi)^2}{4} + \frac{3}{L^2} \cosh(\phi/\sqrt{3})\right)  \text{ .}
\end{align}
In particular, we are looking at long-wavelength excitations in the nodal direction, neglecting possible anisotropy, so that the fields in Eq.~\eqref{eq:scalar} are a function of the radial coordinate $r$ only. This set of equations for $(g_{tt}=-1/g_{rr}, g_{xx}=g_{yy}, A_t, \phi)$ then support a fully analytical black-hole solution \cite{Gubser_2010}, with a non-zero temperature and entropy, and with $A_t(r)$ setting a non-zero density in the dual boundary theory through the radially conserved quantity
$n\equiv \sqrt{-g(r)}~ e^{\phi(r)/\sqrt{3}} F^{rt}(r)$, $g(r)$ being the determinant of the metric. It is important to notice that the definition of the density given here is related to the density of the boundary field theory by the unknown dimensionless prefactor $N_G$ in the action in Eq.~\eqref{eq:dimless_background_action} that we set to unity, as explicitly shown in the appendix. This holds true for all operators and response functions so that, while with a `bottom-up' holographic computation we are able to study the qualitative response of the system, we cannot make a quantitative comparison between the holographic model and the response measured experimentally. However, as we argue in a later section, the introduction of screening effects allows us to fix this constant by matching the experimental plasma frequency, opening up the opportunity for a further, quantitative, test of holographic predictions.  

\begin{figure*}
\begin{subfigure}{.5\textwidth}
    \centering
    \includegraphics[width=\linewidth]{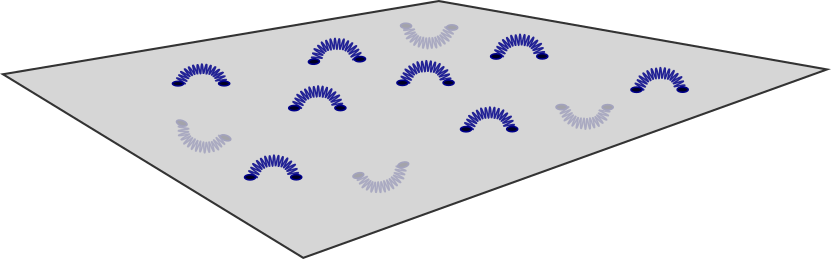}
  \end{subfigure}$\Rightarrow$%
\begin{subfigure}{.35\textwidth}
  \centering
  \includegraphics[width=\linewidth]{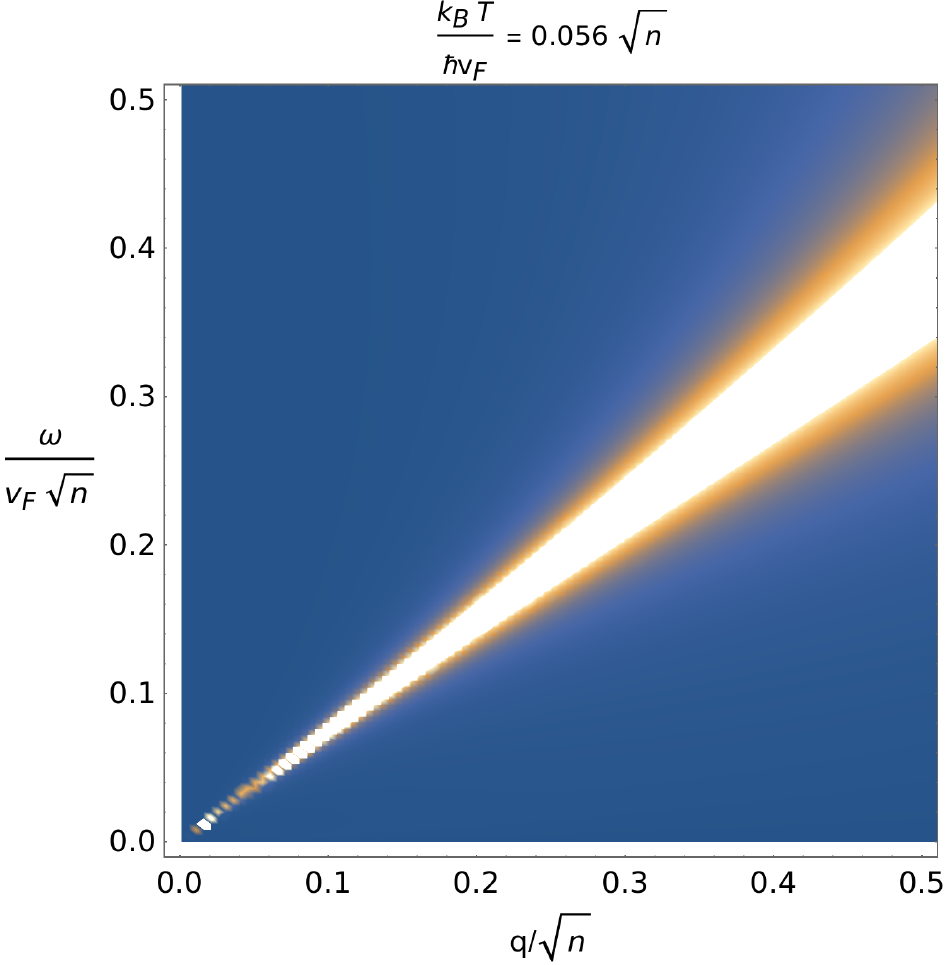}
\end{subfigure}
\caption{\label{fig:neutral_response} (Left) Picture of a single-layer with strong short-range interactions and (right) the imaginary part of the holographic density-density response, $\Pi''(\omega, q)$, at a fixed temperature $k_B T/\hbar v_F \sqrt{n} \simeq 0.038$. We see that the density-density response contains a linear sound mode with $\omega = v_s q$.}
\end{figure*}

In order to also compute the response of the holographic strange metal to small external perturbations, we have to linearize the gravitational equations in Eq.~\eqref{eq:scalar} around the black-hole solution. We obtain in this way a set of coupled equations for the fluctuations $(\delta g_{tt}, \delta g_{tx}, \delta g_{xx}, \delta g_{yy}, \delta A_t, \delta A_x, \delta \phi)$ that can only be solved numerically. According to the holographic dictionary \cite{Hartnoll2009, hartnoll2016holographic, ammon_erdmenger_2015, Zaanen2015}, finding such a solution with infalling-wave boundary conditions at the black-hole horizon, allows us to extract all the retarded Green's function of the system, and hence the density-density response function $\Pi(\omega, \bm q)$, by studying the near-boundary behavior of the field fluctuations. This is a response of a two-dimensional layer with strong in-plane interactions but without screening effects due to the long-range Coulomb interactions, as depicted in Fig. \ref{fig:neutral_response} and evident from the density-density response function that shows a linear sound mode $\omega = v_s q$ instead of the typical plasmon mode expected in the presence of screening $\omega \propto \sqrt{q}$. Here we used rotational invariance to fix $\bm{q} = (q,0)$ without loss of generality. 

\begin{figure*}
\begin{subfigure}{.5\textwidth}
    \centering
    \includegraphics[width=\linewidth]{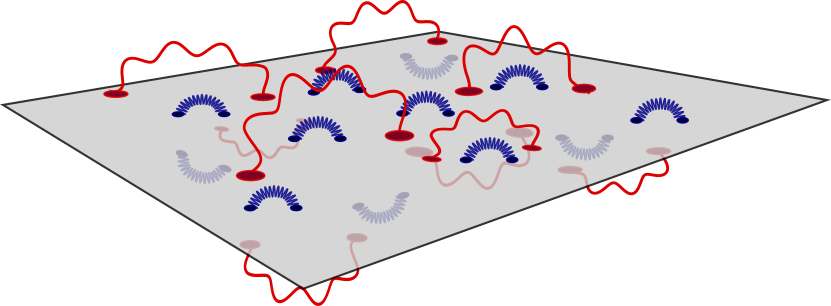}
  \end{subfigure}$\Rightarrow$%
\begin{subfigure}{.35\textwidth}
  \centering
  \includegraphics[width=\linewidth]{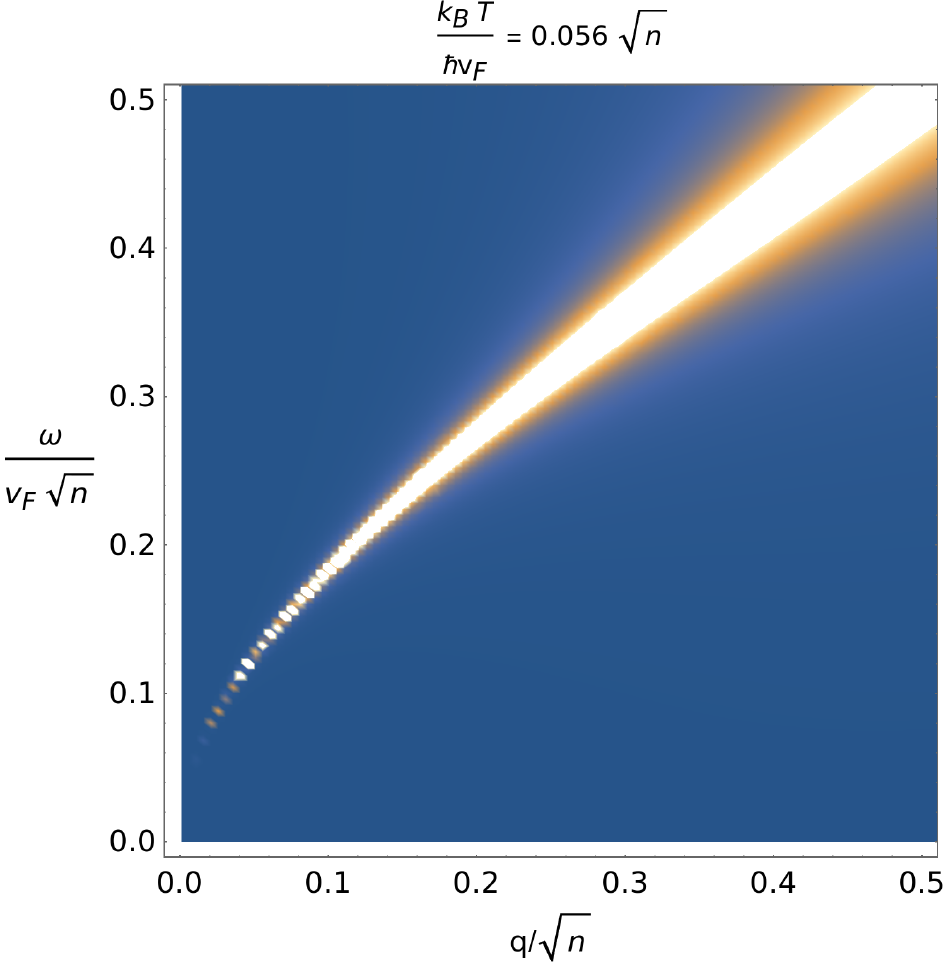}
\end{subfigure}
\caption{\label{fig:plasmon_response} (Left) Picture of a single-layer with in-plane short-range strong interactions (\textcolor{blue}{blue}) and long-range Coulomb interactions (\textcolor{red}{red}) and (right) the holographic density-density response $\Pi''(\omega, q)$ at a fixed temperature $k_B T/\hbar v_F \sqrt{n} \simeq 0.038$. We see that the density-density response contains the plasmon mode with $\omega \propto \sqrt{q}$.}
\end{figure*}

In order to introduce screening effects, we need to couple dynamical photons to the current operator $\avg{J^\mu}$. To do so, as shown in Refs. \cite{Mauri2019, Anomalous_attenuation}, we introduce a boundary ``double-trace'' deformation for the electromagnetic field, i.e., we add to the gravitational action in Eq.~\eqref{eq:background_action} a boundary term  \begin{align}\label{eq:boundary_double_trace}
\begin{split}
    S = &\int \dif r\mathrm{d}^{2} x \dif t  \, \mathcal{L}_\text{gravity}\, +\\ +&\int \mathrm{d}^2 x\dif t\, \int \dif z \left(-\frac{\epsilon}{4} F_{\mu \nu} F^{\mu \nu} + A_\mu \avg{J^\mu} \right) \text{ ,}
 \end{split}
  \end{align}
with $z$ the spatial direction orthogonal to the $x$-$y$ plane, and we used the convention of omitting the tilde on the dimensionless coupling $\tilde\epsilon \equiv \hbar c \epsilon/e^2$, with $\epsilon$ the dielectric constant of the material surrounding the layer and $e$ the electric charge. The addition of this boundary term does not change the linearized equations of motions needed to compute the response function, but it only changes the boundary conditions \cite{Gran2017, Gran2019} for the field fluctuations. In particular, by restricting the current operator to the $x$-$y$ plane where the holographic boundary theory is defined we can see that, with the addition of the boundary double-trace deformation, the density-density response function is related to the `neutral' response function $\Pi(\omega, \bm q)$ by 
\begin{align}\label{eq:2d_response}
    \chi_{\text{2D}}(\omega, \bm q) = \frac{\Pi(\omega, \bm q)}{1 - \frac{\Pi(\omega, \bm q)}{2\epsilon\abs{\bm{q}}}} \equiv \frac{\Pi(\omega, \bm q)}{\epsilon(\omega, \bm q)}\text{ ,}
\end{align}
and we recognize a RPA-like form with a Coulomb potential $V(\bm q) = 1/2 \epsilon \abs{\bm{q}}$, where we neglect retardation effects, i.e., a frequency dependence in the potential, by assuming $v_F \ll c$ (see Ref.~\cite{Mauri2019} for details). 
Two things to keep in mind here are that, contrary to textbook RPA, $\Pi(\omega, \bm q)$ is the density-density correlation function of a strongly interacting system computed from holography, and it hence accounts for the effect of strong interactions as can be seen in Fig. \ref{fig:neutral_response} where we show that $\Pi''$ contains a linear sound mode. Moreover, these interactions are effectively two-dimensional, living only in the plane, and should then be thought of as strong interactions that are short-ranged compared to the inter-layer distance, as  depicted in Fig. \ref{fig:set-up}. On the other hand, the long-range Coulomb interactions described by the addition of the boundary action in Eq.~\eqref{eq:boundary_double_trace} are three-dimensional, as depicted in Fig. \ref{fig:plasmon_response}, where we also show that this induces the expected plasmon dispersion for a single layer.

In order to obtain a simple model of a two-layer cuprate strange metal where the long-range Coulomb interaction is the dominant inter-layer interaction \cite{Hepting2018}, we take two independent copies of a two-dimensional holographic strange metal described by the action in Eq.~\eqref{eq:background_action} and couple them through the three-dimensional boundary double-trace deformation, where the current $J^\mu$ takes the form \begin{align}\label{eq:current_restricted}
  J^\mu(\bm{x}, z) = J^\mu _{(1)}(\bm{x}, z) \delta(z + \ell/2) + J^\mu _{(2)}(\bm{x}, z) \delta(z - \ell/2) \text{ ,}
\end{align}
hence describing two layers laying parallel to the $x$-$y$ plane and separated by a distance $\ell$ along the $z$-axis orthogonal to the $x$-$y$ plane. This again, gives rise to an RPA-like form of the inter-layer density-density response, as we briefly show here. 

First of all, let us remark that, as we want to describe the physics of the strange metal where the linear dispersion has a Fermi velocity $v_F$, we replace from now on the speed of light in the gravitational theory with the Fermi velocity $c \to v_F$. From holography, the effective boundary action for the field fluctuations in Fourier space takes the form
  \begin{align}\label{eq:boundary_effective}
    \frac{1}{2}\int \frac{\dif ^{2} q\dif \omega}{(2 \pi)^3} \sum ^{2}_{i = 1} a ^{i (0)}_\mu(\omega, \bm q) j^\mu_i(-\omega, -\bm q) \text{ ,}
  \end{align}
  where $a_\mu^{i(0)} \equiv \lim_{r \to \infty} \delta{A_\mu^{i}}(r)$ and $j^\mu_i \equiv \delta\avg{J^\mu_i}$.
  Adding the boundary term from Eq.~\eqref{eq:boundary_double_trace} and using the Fourier transform of the current in Eq.~\eqref{eq:current_restricted}, $J(\omega, \bm q, q_z) = J^\mu _{(1)}(\omega, \bm q) e ^{i q_z \ell/2} + J^\mu _{(2)}(\omega, \bm q) e ^{-i q_z \ell/2}$,
  we can integrate out the Maxwell field fluctuations to obtain an effective boundary action for the currents. We define the absolute value of the four-vector $k\equiv \abs{\bm k} = \sqrt{q^2 - \omega^2 v_F^2/c^2}$, where the factor of $v_F^2/c^2$ is due to the above-mentioned fact that the holographic theory is dual to a quantum field theory describing excitations with bare velocity $v_F$ and the dimensionless $\omega$ then contains a factor of $v_F$ while the Coulomb interactions propagate at the speed of light $c$. Then, the three-dimensional double-trace deformation gives a boundary term
  \begin{widetext}
    \begin{align}\label{eq:effective_action_2layer}
    \frac{1}{2}\int \frac{\dif ^{2} q \dif \omega}{(2 \pi )^3}  \left(
      j_1^\mu(\omega, \bm q)\, j_2^\mu(\omega, \bm q)\right)   
    \left(\begin{array}{cc}
       \frac{\eta_{\mu\nu}}{2 \epsilon k}  & \frac{\eta_{\mu\nu} e ^{-k\ell}}{2 \epsilon k}  \\
      \frac{\eta_{\mu\nu} e ^{-k\ell}}{2\epsilon k}  & \frac{\eta_{\mu\nu}}{2 \epsilon k}  \\
    \end{array}\right) 
    \left(\begin{array}{c}
      j_1^\nu(-\omega, -\bm q) \\
      j_2^\nu(-\omega, -\bm q)\\
    \end{array}\right) \text{ .}
  \end{align} 
\end{widetext}
As we are interested in the limit $v_F/c \to 0$, upon a redefinition of $\epsilon c/v_F \to \epsilon$, we then obtain the effective action for the current fluctuations
  \begin{align}\label{eq:effective_action_2layer_nonrel}
    \frac{1}{2}\int \frac{\dif ^{2} q \dif \omega}{(2 \pi ^3)}  \left(
      j_1^\mu\, j_2^\mu\right)   
    \underbrace{\left(\begin{array}{cc}
      {\Pi_1}_{\mu\nu} ^{-1} + \frac{ \eta_{\mu\nu}}{2 \epsilon q}  & \frac{ \eta_{\mu\nu} e ^{-q\ell}}{2 \epsilon q}  \\
      \frac{\eta_{\mu\nu} e ^{-q\ell}}{2 \epsilon q}  &{\Pi_2}_{\mu\nu} ^{-1} + \frac{ \eta_{\mu\nu}}{2 \epsilon q}  \\
    \end{array}\right) }_{\chi^{-1}}
    \left(\begin{array}{c}
      j_1^\nu \\
      j_2^\nu\\
    \end{array}\right) \text{ ,}
  \end{align}  
with $\chi(\omega, q)$ the two-layer response and $q \equiv \abs{\bm q}$. The coupling $\epsilon$ here plays an important role. As mentioned above, in bottom-up computations of holographic theories we can study the behavior of the response function up to an undetermined constant, meaning that we cannot provide a \textit{quantitative} estimate of the magnitude of the effect, as this is only possible from a top-down construction from a known string theory that fixes the prefactor of the action $N_G$. With the introduction of plasmons and hence a known scale $\epsilon$, we can fix this unknown constant to match the plasma dispersion defined by the low-energy pole of Eq.~\eqref{eq:2d_response}.
This allows us a novel check for the holographic framework. Namely, we can further verify if the dual gravitational theory and the proposed plasmon set-up provide also a quantitative agreement with experimental measurements. Looking at it the other way around, studying the plasmon response in cuprates could give insight into the dual string theory needed to match the correct prefactor. We give further details on the relationship between $N_G$ and the plasmon dispersion, and how to translate the dimensionless results from holography into physical units in the appendix.

\section{Coulomb Drag resistivity}\label{sec:coulomb_drag} Our starting point is the drag resistivity within the random-phase approximation that is given in dimensionful units by the expression \cite{PhysRevB.47.4420, PhysRevB.48.8203, PhysRevB.52.14761, DiagramaticApproach}
\begin{widetext}
\vspace{-0.5cm}
 \begin{align}\label{eq:dresistivity}
 \begin{split}
     \rho_{D}=& \frac{\hbar^2}{16 \pi^3 e^2 n_ 1 n_2 k_B T} \int _{-\infty} ^{+\infty} \frac{\dif \omega}{\sinh^2 \left(\hbar \omega / 2 k_B T \right) } \int \dif \bm{q}~ \bm{q}^2 \frac{\text{Im}[\Pi_1(\omega, \bm{q})]}{\abs{\Pi_1(\omega, \bm{q})}^2}\abs{\chi_{12}(\omega, \bm q)}^2 \frac{\text{Im}[\Pi_2(\omega, \bm{q})]}{\abs{\Pi_2(\omega, \bm{q})}^2}\\
     \equiv&  \frac{\hbar^2}{e^2}\int _{-\infty} ^{+\infty} \dif \omega \int_0^{\infty} \dif q\, \mathcal{I}(\omega, q)
     \end{split}\text{ ,}
  \end{align}
 \end{widetext} 
where in the second line we defined the dimensionless integrand $\mathcal{I}(\omega, q)$ and use rotational invariance to choose, without loss of generality, a direction for the in-plane momentum. Here, $n_1$ and $n_2$ are the densities of the active and passive layers, respectively, $\Pi_i(\omega, \bm q)$ are the corresponding intra-layer density-density response functions, and the dynamically screened inter-layer density response is given by the off-diagonal element of $\chi(\omega, q)$ defined through Eq.~\eqref{eq:effective_action_2layer_nonrel}, that is
 \begin{align}\label{eq:inter-layer_response}
    \chi_{12} =
    - \frac{\Pi_1 \Pi_2}{\Pi_1 \Pi_2 \frac{e^2}{\epsilon q} \sinh{(q \ell)} + \left[ \frac{2 q \epsilon}{e^2} - \Pi_1 - \Pi_2 \right] e ^{q \ell} } \text{ .}
  \end{align}
This approximation ignores the effect of vertex corrections. However, as mentioned above, the electron self-energy in cuprates depends mostly on frequency. This feature is also captured in our holographic model where the low-energy theory is described by a `semi-local' quantum critical phase, where the electron self-energy $\hbar \Sigma(\omega, \bm k)$ scales like $\omega^{2 \nu_k}$, with the momentum dependence only entering in the power $\nu_k$  \cite{Gubser_2010, Anantua2013}. We, therefore, expect Eq.~\eqref{eq:dresistivity} to be a valid approximation for the Coulomb drag resistivity for a cuprate two-layer system in the strange-metal regime. 
  
From Eqs.~\eqref{eq:dresistivity} and \eqref{eq:inter-layer_response} we see that the only input we need to compute the drag resistivity are the single-layer response functions $\Pi_i(\omega, \bm q)$. The power of our holographic approach now lies in the fact that it allows us to compute this response in a strongly interacting system. Here in particular, we focus on a regime where the length scale characterizing the strong interaction is much smaller than the lattice scale and, to first order, the behavior of the drag resistivity depends solely on the nature of the inter-layer interactions and not on disorder that would only give higher-order corrections to the drag. This may reveal interesting features of the cuprates, and possibly verify if the proposed holographic model of the strange metal correctly describes this phenomenon.

\section{Results}\label{sec:results} 
\begin{figure}[t]
\includegraphics[width=86mm]{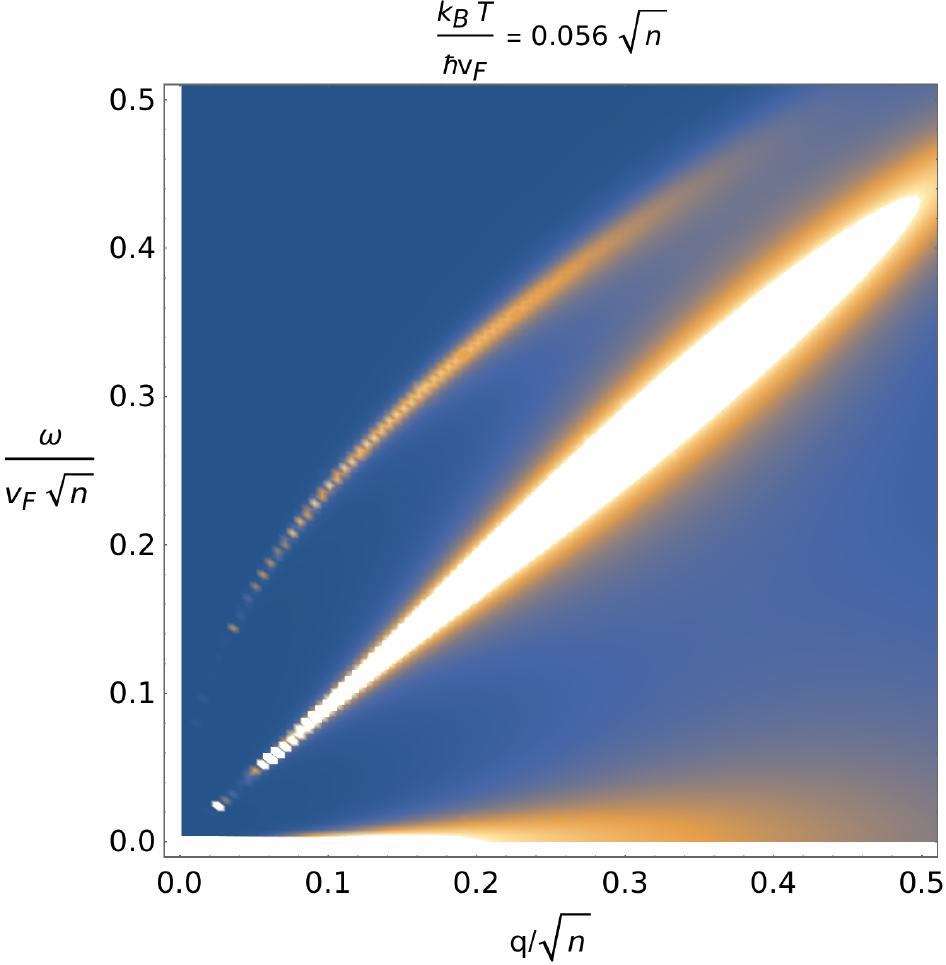}
\caption{\label{fig:density_modes} Typical integrand $\mathcal{I}(\omega, q)$ in Eq.~\eqref{eq:dresistivity}, showing the in-phase and out-of-phase plasmon modes, and the low-energy diffusive mode.}
\end{figure}

We work in each layer with electrons at the same fixed equilibrium density $n$, separately conserved in each layer, and we study the holographic density-density response function that is anticipated to describe the physics of the strange metal near the Fermi momentum, in the regime where the dispersion of the electrons can be linearized. In the low-temperature regime and at energies and momenta much smaller than the Fermi energy and Fermi momentum, respectively, we find that the response function is well approximated by a hydrodynamic model, that takes the form (see Ref.~\cite{Kovtun2012} for details on the derivation) 
\begin{align}\label{eq:hydro_pi}
    \Pi(\omega, q) = \frac{q^2 \left(\omega {\mathcal{D}} + i v_s^2 D_d \chi q^2\right)}{\omega^3 + i \omega^2 q^2 (2 D_s + D_d) - \omega q^2 v_s^2 - i v_s^2 D_d q^4} \text{ .}
\end{align}
Analyzing its pole structure, we see that $\Pi(\omega, k)$ contains a linear sound mode at $ \omega = \pm v_s q - i D_s q^2 + \mathcal{O}(q^4)$, and a diffusive mode at $ \omega = - i D_d q^2 + \mathcal{O}(q^4)$, with $\mathcal{D}$ the Drude weight, and $\chi = -\lim_{q \to 0,\, \omega \to 0} \Pi(\omega, \bm q)$ the (hydrodynamic) compressibility. This form of $\Pi$ implies that the integrand in Eq.~\eqref{eq:dresistivity}, \textit{i.e.} $\mathcal{I}(\omega, q)$, contains, for our symmetric case $\Pi_i \equiv \Pi$ (it is straightforward to generalize the results to layers with different densities and three different surrounding dielectric constants \cite{Badalyan2012}), in addition to a diffusive mode $\omega = - i D_d q^2$, the out-of-phase acoustic plasmon mode and the in-phase optical plasmon characteristic of a two-layer system \cite{Hwang2009, Vazifehshenas2010, Profumo2012, Stauber2012, Badalyan2012}. The plasmon modes have been shown to play a key role in the drag resistivity in two-dimensional electron gases \cite{Flensberg1994, Flensberg1995, Hill1997}. The latter frequencies are
\begin{align}
      \omega =& \pm \sqrt{ v_s ^{2} + \frac{ \mathcal{D}\ell}{2 \epsilon}} q + \mathcal{O}\left( q^2\right) \label{eq:ac_plasmon_mode}\\
      \omega =& \pm \sqrt{\frac{e^2\mathcal{D} q}{\epsilon}} \left( 1 + e^2\frac{v_s^2 -  e^2\mathcal{D} \ell/2 \epsilon} {2 \mathcal{D} \epsilon} q \right) + \mathcal{O}\left( q^2\right) \text{ .} \label{eq:op_plasmon_mode}
\end{align}
All three modes are clearly visible in the numerical result for the integrand in Fig.~\ref{fig:density_modes}. 
These modes thus determine the low-temperature behavior of the drag resistivity. We emphasize here that compared to a purely hydrodynamic approach as in Refs.~\cite{Apostolov2014, Holder2019}, the holographic model gives also a prediction for the coefficients in Eq.~\eqref{eq:hydro_pi}
and for their temperature dependence, allowing us to fully determine the dominant low-temperature behavior of $\rho_D$. In particular, we numerically explored the pole structure of the retarded density-density correlators in the limit of $T \rightarrow 0$ to find that in this low-temperature hydrodynamics \cite{Edalati2010, Karch_2009, Davison2011}, the sound diffusion coefficient equals $D_s = (1/6 \sqrt{3}) k_B T/\hbar n$, whereas the charge diffusion constant obeys \cite{hartnoll2016holographic} $D_d = (4\pi/\sqrt{3}) k_B T/\hbar n$. Notice that these scalings differs from the one found in the Reissner-Nordstr\"om background where there is no linear-in-T resistivity (see, e.g., Ref.~\cite{Zaanen2015}). Both coefficients are fully determined by background (thermodynamic) quantities, as is the zero-temperature `Drude weight' $\mathcal{D} = (1/3^{1/4}) v_F \sqrt{n}/\hbar$, where we use a slight abuse of terminology since the Drude weight for a system with translation invariance usually refers to the strength of the delta function in the zero-frequency limit of the optical conductivity $\sigma(\omega)$. As it is defined here $\mathcal{D}$ is such that $\lim_{\omega \to 0} \sigma(\omega) = \pi e^2 \mathcal{D} \delta(\omega)$. The factor $v_F/3^{1/4}$ appearing in the Drude weight is, as expected, the low-energy speed of sound $v_s$, as we verified numerically finding $v_s \approx 0.76 v_F$. This differs from the $v_F/\sqrt{2}$ of a conformal field theory as conformal invariance is broken in the low-energy limit by the dilaton field. It is also slightly higher than expected from the thermodynamic potential $\Omega$ via $v_s = (-\partial\Omega/\partial \varepsilon)^{1/2} = \sqrt{17/31} v_F$, with the $\Omega$ defined as the gravitational action in Eq.~\eqref{eq:dimless_background_action} evaluated on-shell, and the equilibrium energy density $\varepsilon$ such that it satisfies the thermodynamic relation $-\Omega/V = - \epsilon + s T + \mu n$, with $s$ the entropy density. 
Accordingly, to zeroth order in $T$, $\chi = \mathcal D/v_s^2 \approx 1.32 \sqrt{n}/\hbar v_F$ so that the height of the diffusive peak is determined by the subleading behavior $(\mathcal{D}/v_s^2 - \chi)/2 \approx 2.56 (k_B T)^2/\sqrt{n} (\hbar v_F)^3$, with the proportionality constant obtained numerically. 

\begin{figure}[h]
\includegraphics[width=86mm]{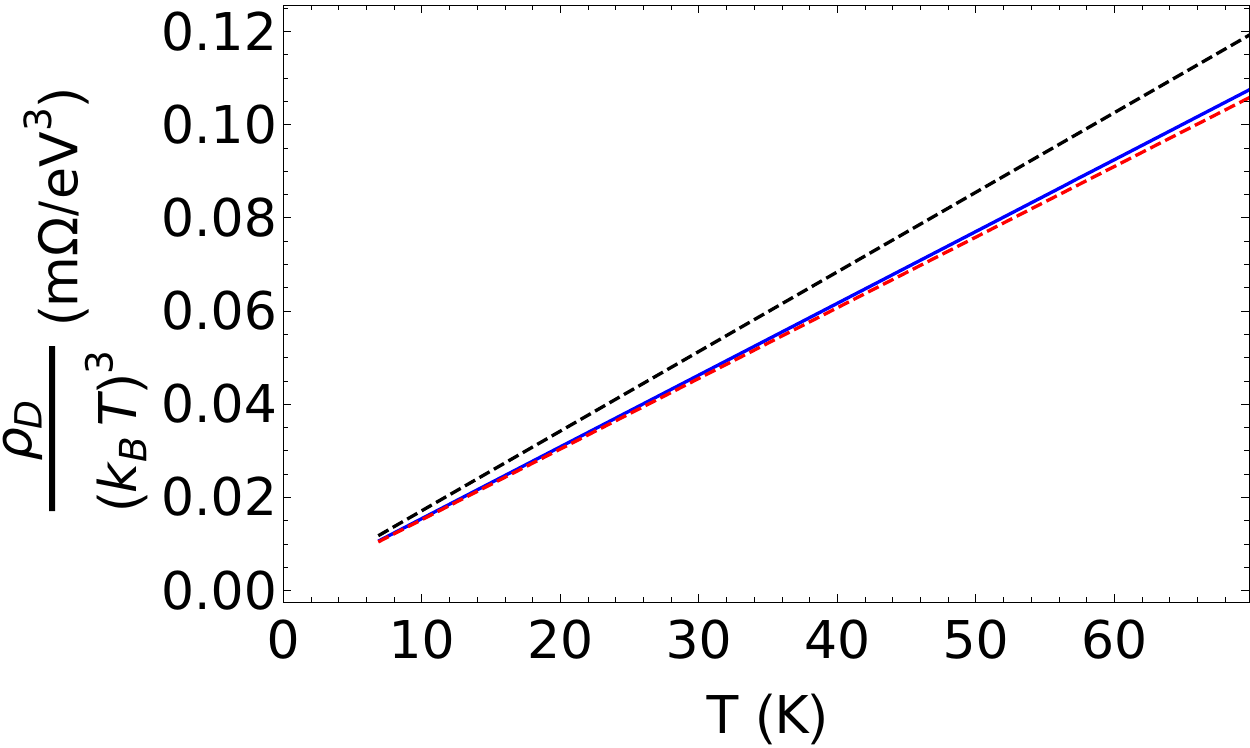}
\caption{\label{fig:low_T_resistivity} 
 Low-temperature scaling of the resistivity in the hydrodynamic approximation (blue line) that shows $\rho_D \propto T^4$ and comparison with the model that only accounts for a diffusive mode as in Eq.~\eqref{eq:hydro_int_diff_approx} (red dashed line). The dashed black line shows the upper-bound estimate in Eq.~\eqref{eq:analytic_approx_rho} that can be computed analytically.}
\end{figure}

The hydrodynamic model in Eq.~\eqref{eq:hydro_pi}, together with the above results, can be integrated to compute the drag resistivity according to Eq.~\eqref{eq:dresistivity}. In general, this can only be done numerically, however, we can look at the dominant behavior in the low-temperature and large $\ell$ limit. In this regime, the dominant contribution to the drag resistivity is determined by the diffusive modes, and we can make the approximation
\begin{align}
    \text{Im}[\Pi] =& -\frac{\omega q^2 D_d (\mathcal{D}/v_s^2 - \chi)}{\omega^2 + D_d^2 q^4} \text{ .}
\end{align}
Moreover, for large $\ell$ the contributions to the drag from the diffusive mode lie at low-energies $\omega \approx D_d q^2$, as the integration range is controlled by a factor $e^{-2q\ell}$ in $\chi_{12}$ from Eq.~\eqref{eq:inter-layer_response}. Hence, $\sinh(\omega/2 T)^{-2} \approx 4 T^2/\omega^2$ and $\Pi \approx -\chi$, and we can then perform the frequency integral to obtain
\begin{align}
\begin{split}
\rho_{D}=& \frac{(\mathcal{D}/v_s^2 - \chi)^2\hbar^2 k_B T}{4 \pi D_d n^2 e^2} \int_0^\infty \dif q~ \frac{e^{-2q\ell}}{\abs{\frac{2 q \epsilon}{e^2} + 2 \chi + \frac{e^2(1 - e^{2 q \ell})}{2 q \epsilon} \chi^2}^2}\\
\approx& \frac{(\mathcal{D}/v_s^2 - \chi)^2\hbar^2 k_B T}{4 \pi D_d n^2 e^2} \int_0^\infty \dif q~ \frac{e^{-2q\ell}}{\abs{ 2 \chi + \frac{e^2(1 - e^{2 q \ell})}{2 q \epsilon} \chi^2}^2} \text{ .}
\end{split}\label{eq:hydro_int_diff_approx}
\end{align}
Given the scaling of the coefficients presented above, this shows that--in the low-temperature limit--$\rho_D(T)\propto T^4$, strikingly different from the Fermi liquid result $\rho_D(T) \propto T^2$ due to thermal broadening of the Fermi surface \cite{PhysRevLett.66.1216, PhysRevB.47.4420, PhysRevB.48.8203}.
In Fig. \ref{fig:low_T_resistivity} we show a plot of $\rho_D/(k_B T)^3$ at low temperature for $\ell = 75 \text{\AA}$ with the numerical integration of the full hydrodynamic model from Eq.~\eqref{eq:hydro_pi} (blue line) and the diffusive-mode expression in Eq.~\eqref{eq:hydro_int_diff_approx}, that shows the validity of the approximation in this regime. The dashed black line gives an upper-bound analytical estimate of the resistivity given by 
\begin{align}\label{eq:analytic_approx_rho}
  \begin{split}
    \rho_D \lesssim& \frac{(\mathcal{D}/v_s^2 - \chi)^2\hbar^2 k_B T}{4 \pi D_d n^2 e^2} \int_0^\infty \dif q~ \frac{e^{-2q\ell}}{\abs{\frac{e^2(1 - e^{2 q \ell})}{2 q \epsilon} \chi^2}^2}\\
    =& \frac{3 \zeta(3) \hbar^2}{8\pi n^2 e^2 \alpha^4}\frac{(\mathcal{D}/v_s^2 - \chi)^2 k_B T}{D_d \chi^4 \ell^4}
     \propto \frac{T^4}{\ell^4}  \text{ .}
    \end{split}
\end{align}

\begin{figure}[h]
\includegraphics[width=86mm]{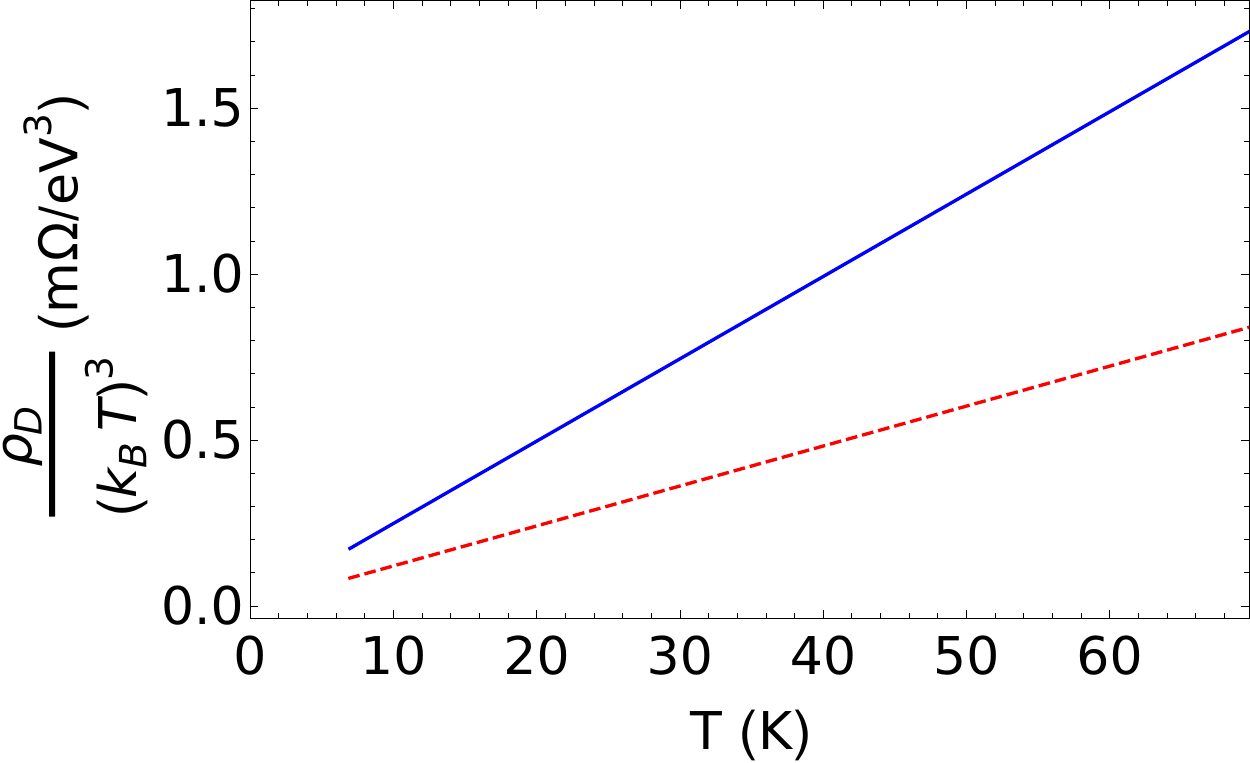}
\caption{\label{fig:low_T_resistivity_cuprate} 
 Low-temperature scaling of the resistivity in the hydrodynamic approximation (blue line) that shows $\rho_D \propto T^4$. Here we used the parameters of the cuprate modeled in this paper, that we introduced above. This shows that in this regime the resistivity presents the scaling expected from the diffusive mode (dashed red line), even though considering the diffusive mode alone underestimates its magnitude.}
\end{figure}

For the realistic system we want to model,  we use the density $n = 6.25 \times 10^{14}\, \text{cm}^{-2}$, the Fermi velocity obeying $\hbar v_F = 3\, \text{eV\AA}$ as typical in the cuprates, and the inter-layer spacing $\ell = 6.2\, \text{\AA}$, that is of the order of the effective inter-layer distance of LCCO planes used in the experiment on plasmons by Hepting \textit{et al.} \cite{Hepting2018}, and we assumed a dielectric constant of the insulating material between the layers of $\epsilon = 2 \epsilon_0$. Further, we are able to give a quantitative result thanks to the above-mentioned matching of the prefactor $N_G$, otherwise undetermined in a bottom-up approach, with the experimentally determined plasma dispersion (see appendix for the details).
 For this smaller value of the interlayer distance, the full hydrodynamic result is considerably larger than expected from the response function with only a diffusive mode. Nonetheless, it shows the same $T^4$ dependence at small temperatures, where the contribution of the plasmon modes is still irrelevant, as shown in Fig.~\ref{fig:low_T_resistivity_cuprate}.
 It is important to notice that the temperature dependence of the dissipative coefficients $D_d$ and $D_s$, that ultimately leads to the $T^4$ scaling of the drag resistivity, is characteristic of the particular holographic model used. 
 However, the linear dependence on $T$ of these coefficients comes from the linear-in-$T$ resistivity, which is a fundamental requirement of any strange-metal model. 
 
\begin{figure}[h]
\includegraphics[width=86mm]{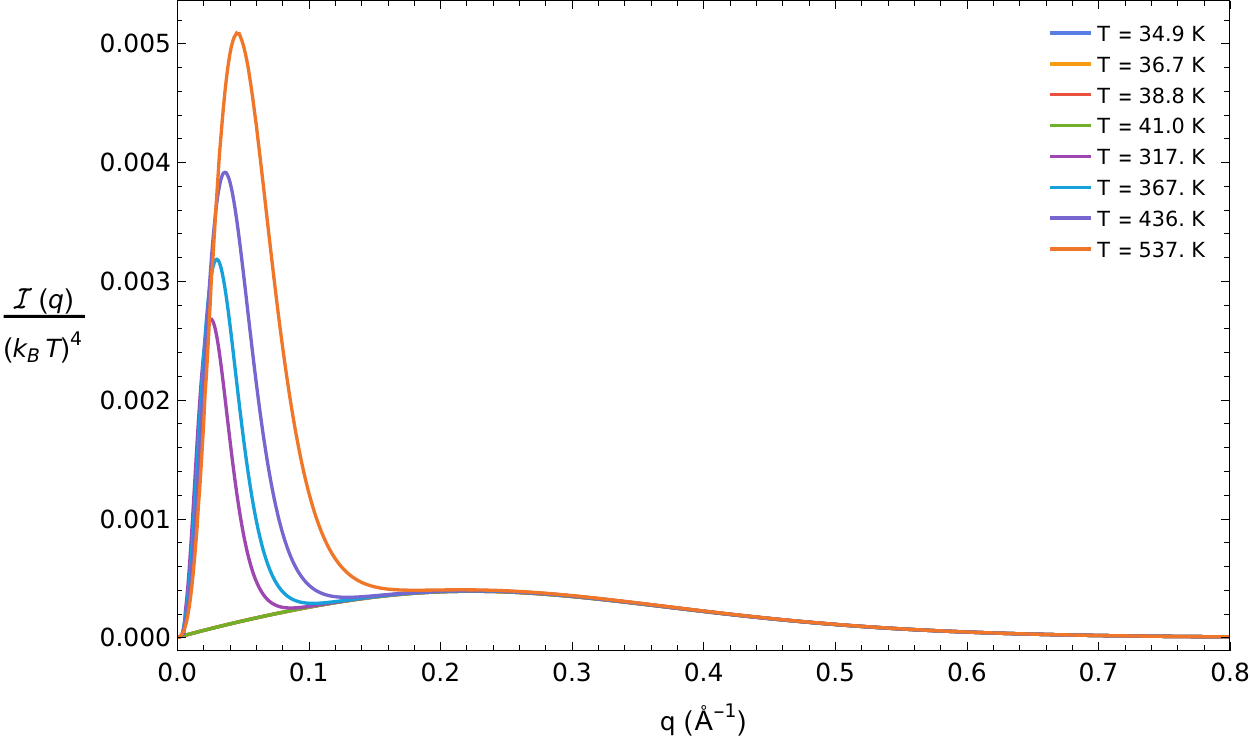}
\caption{\label{fig:low_T_integrand} 
 Result of the integral over frequency in Eq.~\eqref{eq:dresistivity} for the drag resistivity as a function of momenta rescaled with $(k_B T)^4$. We see that at low temperatures the contribution from the plasmon modes is negligible and the resistivity scale as $\rho_D \propto T^4$, while as the temperature is raised the contribution to the drag of the plasmon modes becomes more and more dominant.}
\end{figure}

\begin{figure}[h]
\includegraphics[width=86mm]{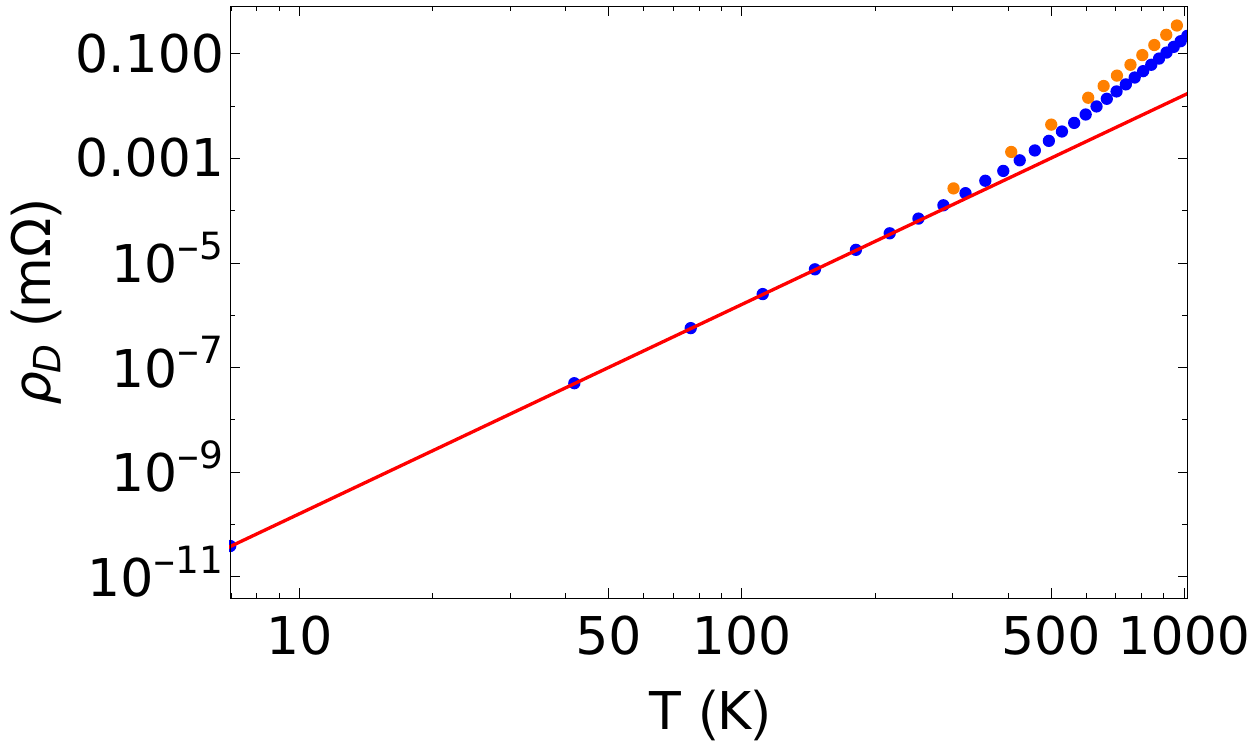}
\caption{\label{fig:drag_resistivity_loglog} 
 Log-log plot of the drag resistivity in the hydrodynamic approximation (blue dots) where we can clearly see the deviation from the $T^4$ behavior (represented by the red line) above room temperatures. We also show the results for the drag resistivity obtained from the holographic model (orange dots), as explained below.}
\end{figure}

 While the $T^4$ scaling controls the low-temperature behavior, at room temperature and higher temperatures, the above approximation breaks down. In this regime, the contribution of both the optical and the acoustic plasmon modes cannot be neglected and the drag resistivity grows faster with temperature. This is shown in Figs.~\ref{fig:low_T_integrand} for the hydrodynamic approximation, where we plot for different temperatures the result of the frequency integral in Eq.~\eqref{eq:dresistivity}, i.e. $\mathcal{I}(q) \equiv \int \dif \omega\, \mathcal{I}(\omega, q)$, rescaled with a factor $(k_B T)^4$. We see that at low temperatures the contribution from the plasmon modes is negligible and the integrand scales as expected from the argument presented above. As the temperature is increased the function develops a peak at low momenta, which is due to the contribution of the plasmon modes and does not scale with $T^4$. The effect of the plasmons and the change of the scaling behavior can also be seen in the $\log$-$\log$ plot in Fig.~\ref{fig:drag_resistivity_loglog}. With the cuprate we are modeling in this paper, characterized by the quantities specified above, the system at higher temperatures enters a regime where both plasmon modes are relevant and we cannot simply neglect the contribution of one or the other. For this reason, together with the fact that at the higher values of $T$ the exponential factor $\sinh(\omega/2T)^{-2}$ in Eq.~\eqref{eq:dresistivity} starts to dominate, the scaling of the resistivity in this regime cannot be described by a simple power law.
 
 
\begin{figure}[h]
\includegraphics[width=86mm]{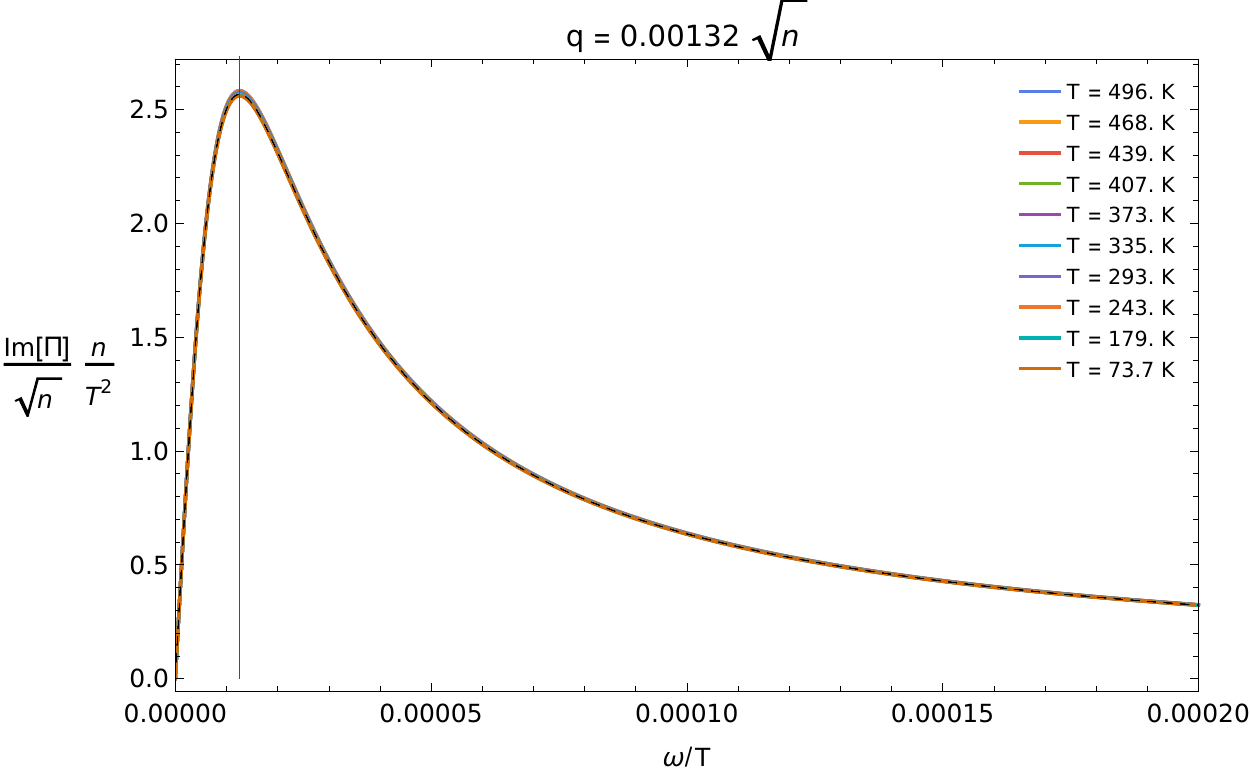}
\caption{\label{fig:collapsed_diff_0_001} 
 Data collapse of the imaginary part of the density-density response function at low energies - that determines the low-temperature drag resistivity - for several temperatures, together with the behavior expected from the hydrodynamic model (dashed black line), for one value of momentum relevant to the low-temperature resistivity $q = 0.00132 \sqrt{n}$}
\end{figure}

\begin{figure}[h]
\includegraphics[width=86mm]{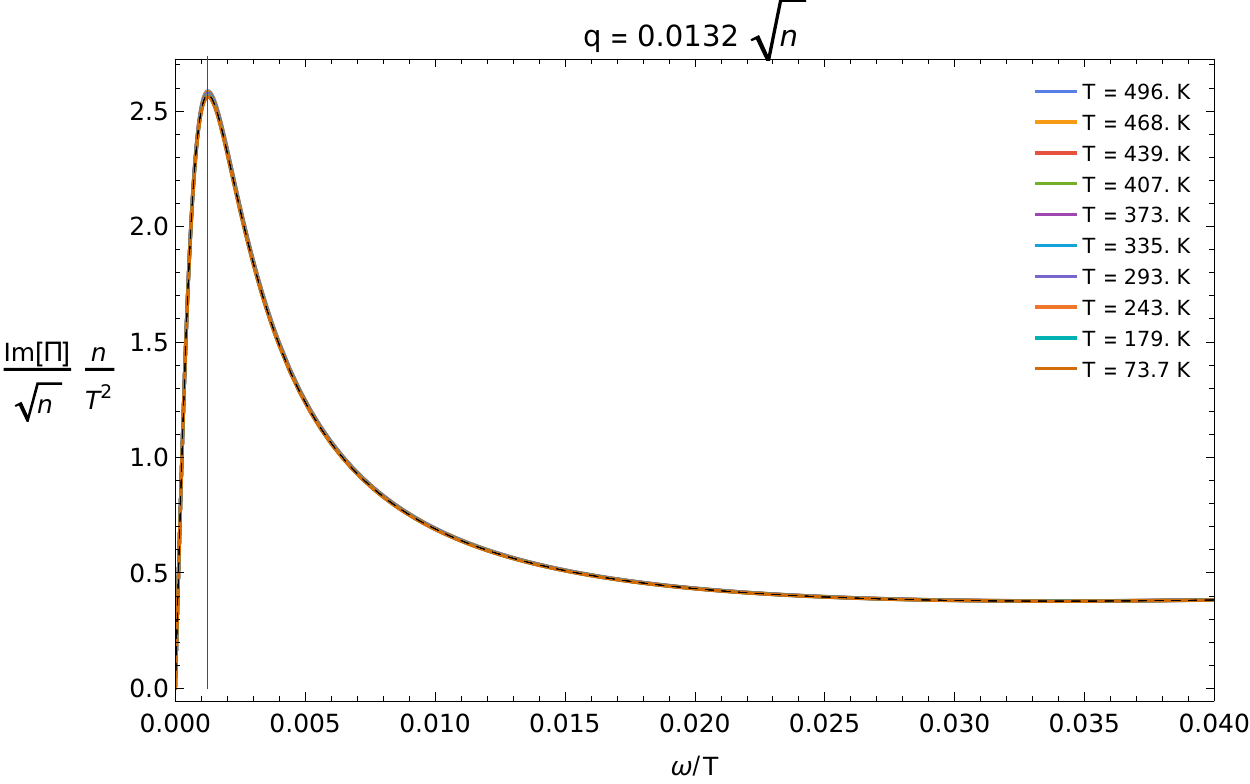}
\caption{\label{fig:collapsed_diff_0_01} 
 Data collapse of the imaginary part of the density-density response function at low energies, together with the behavior expected from the hydrodynamic model (dashed black line), at $q = 0.0132 \sqrt{n}$}
\end{figure}

 As of now, we looked at a hydrodynamic model, where the input from holography comes in the form of a prediction for the temperature dependence of the coefficients in the model. While we expect a deviation between the hydrodynamic model and the holographic response at larger temperatures, we show below that in the low-temperature limit we do not expect such a deviation to change the drag resistivity significantly. The low-energy physics described by the holographic EMD model is, in fact, not fully described by hydrodynamics, as it contains a quantum critical contribution characterized by a dynamical critical exponent $z = \infty$. In particular, this implies that there is a quantum critical contribution to the low-energy spectral function, with $\hbar\omega \ll k_B T$, that scales with temperature as $\Pi \propto T^{2\nu_q}$, with $\nu_q$ a momentum dependent exponent \cite{Hartnoll:2011pp}. One might then wonder if this low-energy scaling could affect the low-temperature scaling of the drag resistivity. However, in the holographic model used in this paper, we found, for the lowest temperatures accessible to our numerics where the drag resistivity is determined mostly by the low-frequency behavior of $\Pi$, that the temperature dependence of the modes in the integrand in Eq.~\eqref{eq:dresistivity} is well described by the hydrodynamic model, and corrections due to the momentum-dependent temperature scaling from the quantum critical sector can be neglected. This is pictured in Fig.~\ref{fig:collapsed_diff_0_001} and \ref{fig:collapsed_diff_0_01} where we show, for some fixed momenta in our range of interest, that the diffusive mode in $\text{Im}[\Pi]$ computed from holography scales as expected from the hydrodynamic model (dashed red line in the plots), with deviations that are only of higher order in temperature. It might be nonetheless an interesting point for future studies to check if there might be a setting where the quantum critical scaling becomes relevant to the drag resistivity even at low temperatures and if it could then be possible to observe it experimentally. 
 
 
\begin{figure}[h]
\includegraphics[width=86mm]{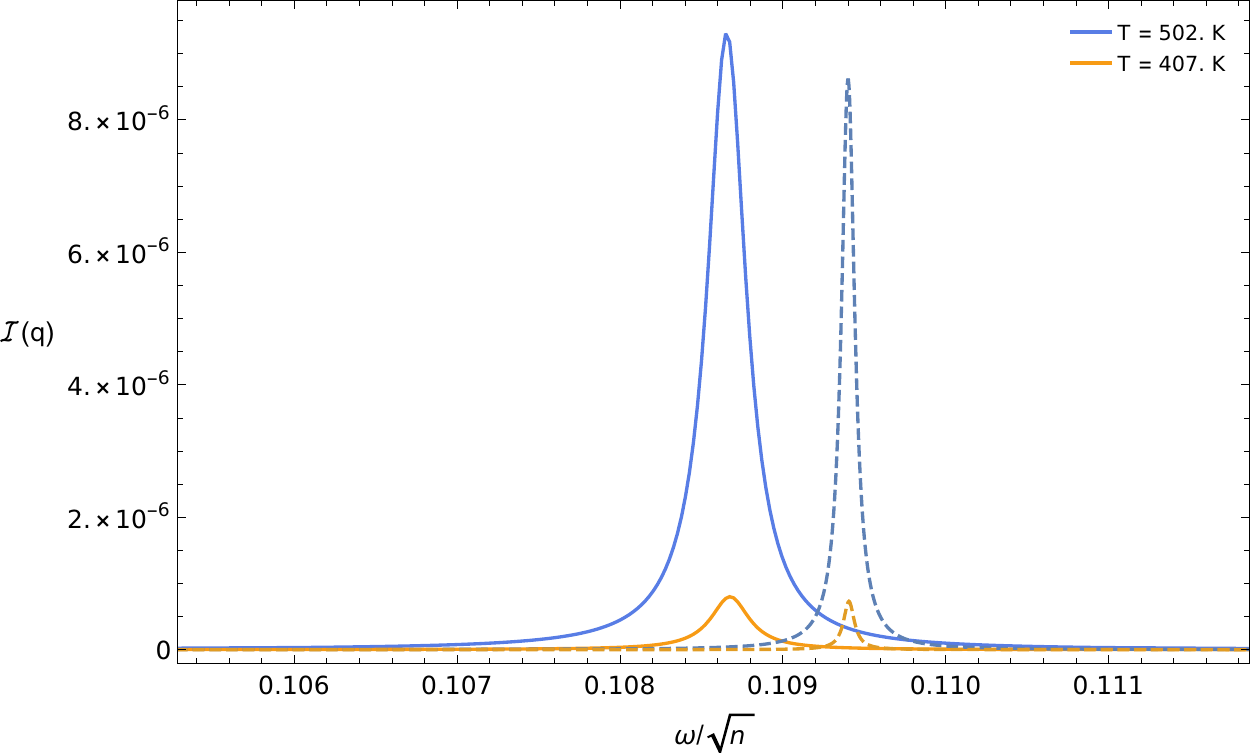}
\caption{\label{fig:plasmon_comparison} 
 Comparison of the integrand function $\mathcal{I}(\omega, q)$ as defined in Eq.~\eqref{eq:dresistivity} at fixed $q = 0.066 \sqrt{n}$ and high temperatures. We can see that the holographic result (solid lines) shows an enhancement of the plasmon contribution compared to the one expected from the hydrodynamic model (dashed lines).}
\end{figure}

\begin{figure}[h]
\includegraphics[width=86mm]{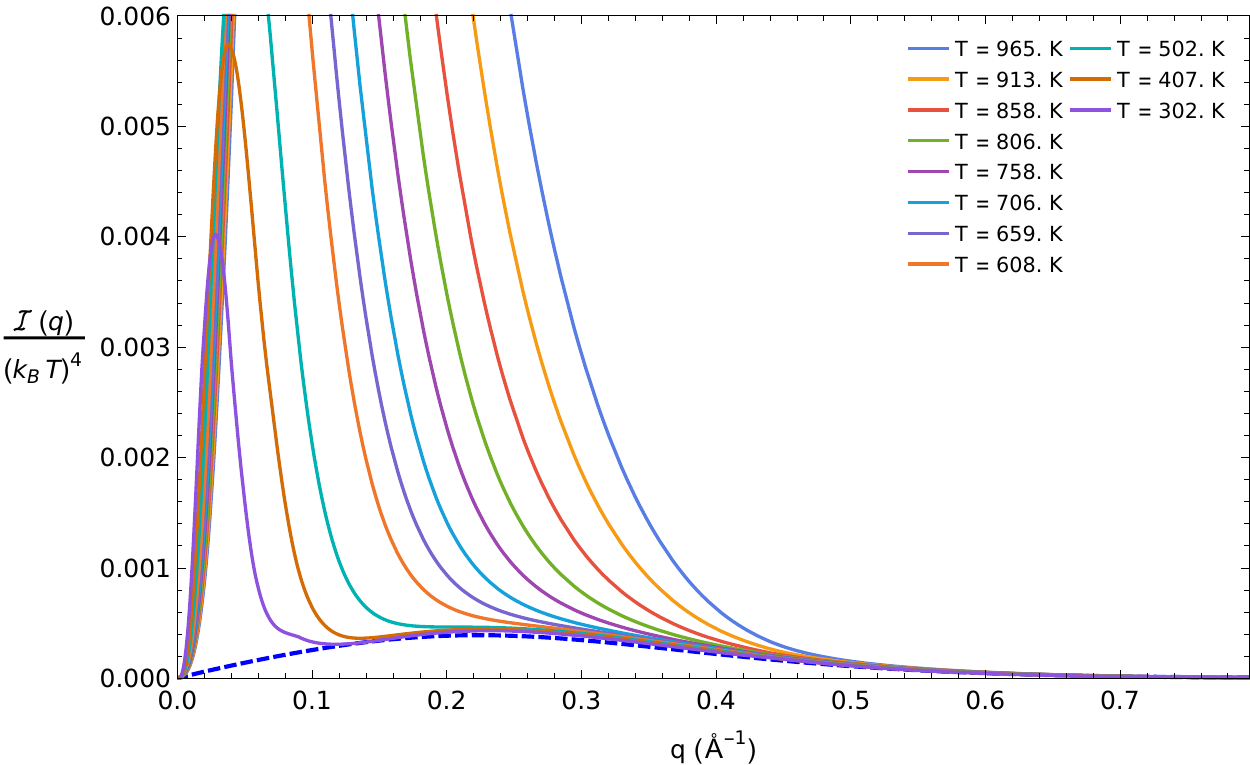}
\caption{\label{fig:rescaled_num_integrand} 
 Result of the integral over frequency in Eq.~\eqref{eq:dresistivity} for the drag resistivity as a function of momenta rescaled with $(k_B T)^4$ for the holographic model (solid lines). We see that the contribution $\propto T^4$ that dominates at low temperatures is well captured by hydrodynamics (dashed blue line).}
\end{figure}

\begin{figure}[h]
\includegraphics[width=86mm]{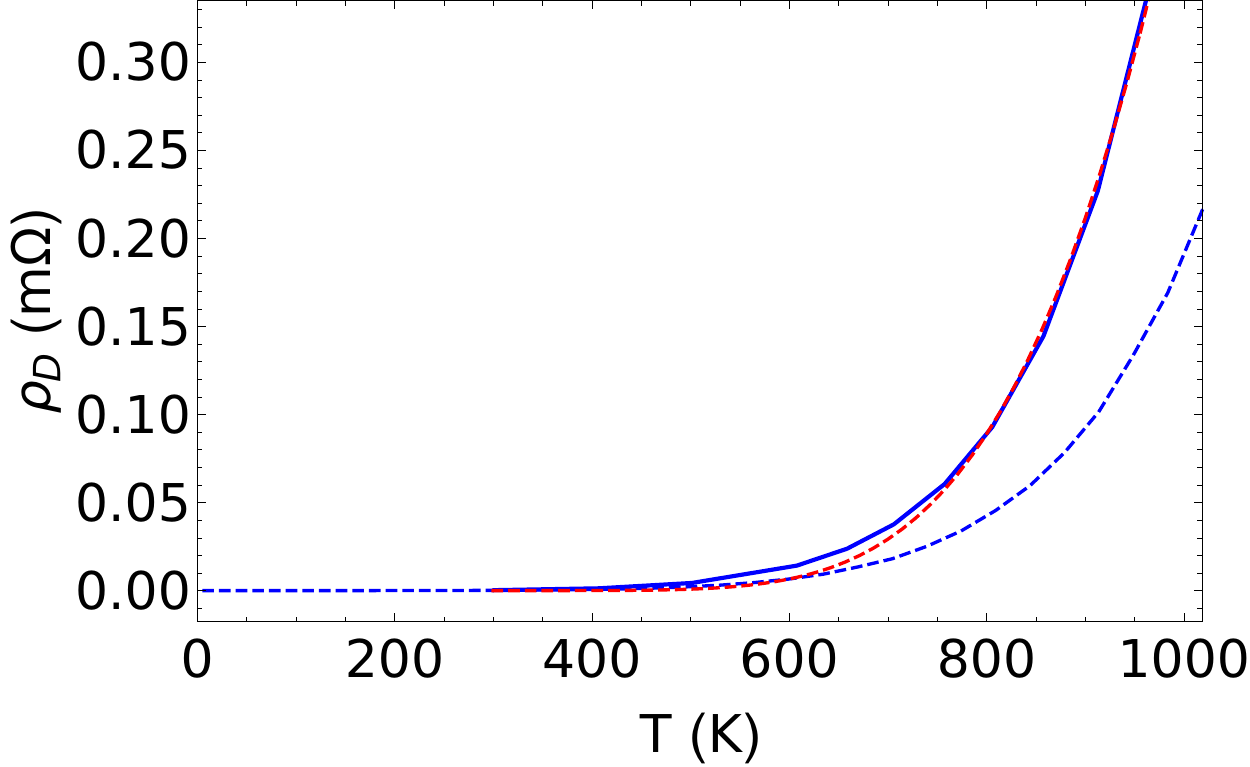}
\caption{\label{fig:drag_resistivity} 
 Coulomb drag behavior at large temperatures, where its behavior is determined by the plasmons, departing from the $\rho_D \propto T^4$ scaling of the low-temperature limit. Here we show the results of the holographic resistivity (solid blue line) and compare it with the hydrodynamic model (dashed blue line), showing an enhancement at high temperatures. In particular at temperature above approximately $800 \text{K}$ the drag resistivity enters a regime where $\rho_D(T) \propto e^{-\Delta/T}$ (dashed red line), where we found $\Delta \approx 6240 \text{K}$}
\end{figure}

 At higher temperatures, however, the relevant energy range for the computation of the drag resistivity becomes large enough that the deviation of the holographic solution from its approximation with the hydrodynamic model cannot be neglected, and we hence need a full numerical solution of the holographic model to compute the resistivity integral. In particular, there is an enhancement of the plasmon modes contribution in the holographic solution, due to higher-order corrections neglected in the hydrodynamic model, as well as the quantum critical contribution. A distinction between the roles of these two contributions goes beyond the scope of this paper, but it can be an interesting problem to address in future studies. The greater contribution from the plasmon mode is shown for example in Fig.~\ref{fig:plasmon_comparison}, where we plot the acoustic plasmon peak in the integrand $\mathcal{I}(\omega, q)$ at a fixed value of $q = 0.066 \sqrt{n}$ for both the holographic model and its hydrodynamic approximation. As highlighted in Fig.~\ref{fig:rescaled_num_integrand}, where we look at the integral over frequency rescaled with temperature, $\mathcal{I}(q)/(k_B T)^4$,  the $T^4$ scaling that governs the low-temperature resistivity follows closely the hydrodynamic approximation (dashed blue line). Hence we expect the holographic drag resistivity to deviate more significantly from the hydrodynamic result at higher temperatures, as the contribution from the plasmon modes becomes more and more dominant. This is shown in Fig.~\ref{fig:drag_resistivity} where we present the result for $\rho_D(T)$ up to high temperatures in holography (solid blue line) and hydrodynamics (dashed blue line). 
 %The logarithmic plot in Fig.~\ref{fig:drag_resistivity_loglog}, emphasize the departure from the $T^4$ scaling above room temperatures. 
 In particular, we see that in the holographic model just above room temperature there is a regime with a contribution to the temperature dependence coming from the different lifetimes of plasmon excitations at different temperatures. On the other hand, as we increase the temperature further, we enter a regime where the behavior of the drag resistivity is determined by the $\sinh(\omega/2T)^{-2}$ factor, hence  $\rho_D(T) \propto e^{-\Delta/T}$, (dashed red curve in the plot). We found $\Delta \approx 6240 \text{K}$, consistent with the acoustic plasmon energy.    
 %We would also like to highlight that, due to the introduction of the Coulomb interactions into the holographic model, we are able to obtain also the magnitude of the drag effect. In a bottom-up approach to holography we are only able to compute response functions up to an unknown factor related to (dimensionless) Newton's constant $16 \pi G\hbar/c^3L^2$. The introduction of the Coulomb interaction, and hence the plasma frequency, allows us to match this unknown constant with the experimentally observed plasma frequency and uniquely fix the magnitude of the density response.

\section{Conclusions and outlook} We have here studied by holography the Coulomb drag between two strange metals and have shown how to determine the drag resistivity in such a strongly interacting system. We performed the computation within a particular model for which an analytical background solution exists and showed the important role of the plasmon modes in the drag resistivity at easily accessible temperatures. We found that the temperature scaling of the drag resistivity is governed at low temperatures by the hydrodynamic modes, whose scaling is fixed by the background solution of our holographic theory and is related to the fact that the model describes a strongly interacting system with linear-in-temperature resistivity.
In particular, we showed that in the low-temperature limit the contribution from the plasmon modes is negligible and the scaling of drag resistivity follows from the scaling of the diffusion constants in the holographic density response function $\Pi(\omega, q)$, implying $\rho_D \propto T^4$ contrary to the $T^2$ behavior at low temperature in a Fermi liquid where the drag is governed by thermal broadening of the Fermi surface.
As the temperature is raised above room temperature we found that the drag resistivity departs from the $T^4$ scaling due to the contribution of both the in-phase and out-of-phase plasmon modes. Here the holographic solution shows an enhancement of the drag resistivity compared to the hydrodynamic approximation, which neglects the quantum critical contribution.  

The holographic model used captures various qualitative features of the low-energy properties of the strange metal. However, we also have to keep in mind that it describes a system with a very different ultra-violet behavior compared to a laboratory cuprate material as the electronic dispersion is implicitly linearized and no lattice bandstructure is involved in the calculation. While this does not affect the emergent $T^4$ scaling behavior expected at low temperatures, it may affect the quantitative estimate of the drag resistivity somewhat. An important matter of future studies is the thermodynamics of the strange metal, as an analytical understanding of the observed values for the speed of sound and susceptibility is still lacking, as well as an analytical understanding of the role of the quantum critical sector in the drag resistivity. We, nevertheless, hope that the theory presented here may stimulate further experiments on drag transport to test holographic quantum matter in general and strange metals in particular.

{\it Acknowledgments.} --- We are grateful to Alberto Cortijo for helpful suggestions and discussions. This work is supported by the Stichting voor Fundamenteel Onderzoek der Materie (FOM) and is part of the D-ITP consortium, a program of the Netherlands Organization for Scientific
Research (NWO) that is funded by the Dutch Ministry
of Education, Culture and Science (OCW).

\appendix

\section{Prefactors and physical quantities}

Here we explicitly show the relations between the dimensionless units used in the main text and the dimension-full variables, taking particular care in considering all the prefactors that have been set to unity in the main text for notational convenience. This makes clear the choice of parameters used to match the expected plasma frequency. 
In this section, we reintroduce the tilde notation for dimensionless variables, e.g. $\tilde n = L^2 n$. Remember that with our definition of dimensionless quantities we have that lengths are measured in units of $L$, the AdS radius, energies in units of $\hbar v_F/L$, and the electric charge in units of $\hbar/L g_F$. The action in these units then takes the form of Eq.~\eqref{eq:dimless_background_action}.
%, that we rewrite here for convenience. 
%\begin{align}
% \begin{split}
%    \tilde S_{\text{EMD}} = \frac{c^3 L^2}{16\pi \hbar G} \int& \dif \tilde r\, \dif \tilde t\, \dif^2 \tilde{x}\sqrt{-g} \left[R - \frac{(\partial_\mu\phi)^2}{2} \right. \\&+\left. 6 \cosh\left(\frac{\phi}{\sqrt{3}}\right) - \frac{e^{\phi/\sqrt{3}}}{4} \tilde{F}_{\mu\nu}^2 \right] \text{ .}
% \end{split}
%\end{align}
In order to further simplify the notation, we redefined the dimensionless action to set the prefactor $c^3 L^2/16\pi \hbar G \equiv N_G$ to 1, which led to the definition of the dimensionless density as in the main text. However, the expectation values of boundary operators and, hence, boundary response functions are proportional to this unknown coefficient. To give a concrete example, using the holographic dictionary \cite{Zaanen2015} that defines the chemical potential of the boundary theory as $\lim_{r \to \infty} \tilde A_0 \equiv \tilde \mu/\tilde e$, with $\tilde e$ the dimensionless electric charge, the corresponding density of the dual field theory can be computed by studying the (renormalized) boundary action to be 
\begin{align}\label{eq:density_functional_of_mu}
    \tilde{n} = L^2 n = \frac{1}{\sqrt{3}}\frac{N_G}{\tilde e} \left(\frac{\tilde\mu}{\tilde{e}}\right)^2\sqrt{1 + \frac{\tilde e^2 \tilde T^2}{3 \tilde\mu^2}} \text{ .}
\end{align}
From now on, we redefine $\tilde \mu/\tilde e \to \tilde \mu$ and $\tilde s_0 = N_G/\tilde e$, in order to get rid of the dimensionless charge. 

In bottom-up holography, the unknown constant $\tilde s_0$ is usually neglected and we are simply concerned with the qualitative behavior of the response function. By introducing the double-trace deformation that leads to plasmon modes in the system, we introduce a known scale given by the plasma frequency that we can use to uniquely determine the constant $\tilde s_0$. Looking at a single-layer cuprate, we have that the plasma frequency is defined by the solution of
\begin{align}\label{eq:2D_plasmon_equation}
    \Pi^{-1}(\omega, q) - \frac{e^2}{\epsilon} \frac{1}{2 q} = 0 \text{ ,}
\end{align}
with $\epsilon$ the dielectric constant of the insulating material surrounding the layer. At low momenta, we found that the holographic model takes the hydrodynamic form in Eq.~\eqref{eq:hydro_pi}, which allows us to express the plasma frequency in terms of the Drude weight. In particular, we found that in our EMD theory the Drude weight in the units used in the main text is
\begin{align}\label{eq:dimless_D}
   \mathcal{\tilde D}(n, T) \equiv \frac{\hbar L}{v_F} \mathcal{D} = \frac{\tilde s_0}{\sqrt{3}} \tilde \mu(n, T) \text{ ,}
\end{align}
where we remind the reader that the dimensionful $\mathcal{D}$ is defined such that $\lim_{\omega \to 0} \sigma(\omega) = \pi e^2 \mathcal D \delta(\omega)$, and we made it explicit that $\tilde\mu$ is a function of the background density and temperature. In fact, for the numerical computation, it is often convenient to work at a fixed chemical potential and work with quantities that are rescaled by it, like $\tilde \omega/\tilde\mu,\, \tilde \Pi/\tilde\mu$, etc. Ultimately however, we work in a system at a fixed density and temperature, so that for every value of $T$ and $n$, $\tilde \mu$ is determined according to Eq.~\eqref{eq:density_functional_of_mu}, by $\tilde\mu(\tilde n, \tilde T) = \sqrt{\sqrt{108 \tilde s_0^2 \tilde n^2 + \tilde T^2} - \tilde T^4}/\sqrt{6}$. At $T = 0$ it reduces to the expressions presented in the main text in terms of $\tilde n$, since for $\tilde s_0 = 1$ we have $\tilde\mu = 3^{1/4} \sqrt{\tilde n}$. There we were only interested in the leading temperature behavior and we could neglect the above subleading temperature correction from the chemical potential.
From Eq.~\eqref{eq:2D_plasmon_equation}, we find that the plasma dispersion at low momenta is given by 
\begin{align}
    \hbar \omega_\text{pl}(q) = \hbar \sqrt{\frac{\mathcal{D}}{2\pi \epsilon}}\sqrt{q} = \sqrt{\frac{\tilde s_0 \mu(n, T)}{2\sqrt{3} \tilde\epsilon}}\sqrt{\hbar v_F q} \text{ ,}
\end{align}
with $\tilde \epsilon \equiv \hbar v_F \epsilon/e^2$. Hence, for every fixed value of the density and temperature, we can determine $\tilde s_0$ from the plasmon dispersion. 

In this paper we aimed at modeling the cuprate layers as in Ref.~\cite{Hepting2018}, where they look at an (effectively) infinite stack of layers. In such a system they find that the plasma frequency for in-phase plasma excitations (i.e., when the out of plane momentum $q_z = 2\pi N$, $N \in \mathbb{N}$) obeys $\hbar \omega_\text{pl}(q_z = 0) = 1.17 \text{ eV}$. The equivalent expression for the plasma frequency in terms of the quantities defining our holographic model is
\begin{align}
    \hbar \omega_\text{pl} = \hbar \sqrt{\frac{\mathcal{D}}{\pi \epsilon}}\frac{1}{\sqrt{\ell}} = \sqrt{\frac{\tilde s_0 \mu(n, T)}{\sqrt{3} \tilde\epsilon}}\sqrt{\frac{\hbar v_F}{\ell}} = 1.17 \text{ eV} \text{ .}
\end{align}
(See Ref.~\cite{Mauri2019} for details of the computation). Notice that here $\mu(n, T)$ is the boundary chemical potential in dimensionful units. 
This is what we use to fix the scale $\tilde s_0$ and to translate the quantities obtained from the holographic computation in terms of  SI units. In particular, at $T = 0$ we have that 
\begin{align}
    \sqrt{\tilde s_0} = 3^{1/4} \frac{\ell}{\sqrt{n}}\frac{(\hbar \omega_\text{pl})^2}{\hbar v_F} \frac{\epsilon}{e^2} \text{ .}
\end{align}
Once we determined $\tilde s_0$ all the results obtain from the numerical computation can then be expressed in terms of SI units by fixing a value for $T$, $n$, $\ell$, and $v_F$. Calling $P_D$ the numerical integral for the drag in rescaled dimensionless units where $N_G = 1$, we have that the drag resistivity in Ohm is then related to $P_D$ by 
\begin{align}
    \rho_D = \frac{\hbar}{e^2 \tilde s_0^2} P_D\text{ .}
\end{align}
Explicitely, in this paper we used $n = (0.25 \text\AA^{-1})^2 = 6.25 \times 10^{14} \text{cm}^{-2}$, $\hbar v_F = 3 \text{eV}\text\AA$, $\ell = 6.2 \text\AA$, $\epsilon = 2 \epsilon_0 = 2 \times 55.26 10^{-4} e^2 \text{eV}^{-1}\text\AA^{-1}$.
Fixing the in phase infinite-layer plasma frequency to be independent of temperature and equal to $\hbar \omega_\text{pl} = 1.17 \text{eV}$ we obtain a (temperature-dependent) $\tilde s_0$ and a value for the holographic scale $\mu$ shown in Fig.~\ref{fig:s0_mu_vs_T}. This last scale is what we then use to set all of the dimensionless quantities used in the numerical computation. 

\begin{widetext}
\begin{center}
      \begin{figure}[h]
  \begin{subfigure}{.48\textwidth}
    \centering
    \includegraphics[width=\linewidth]{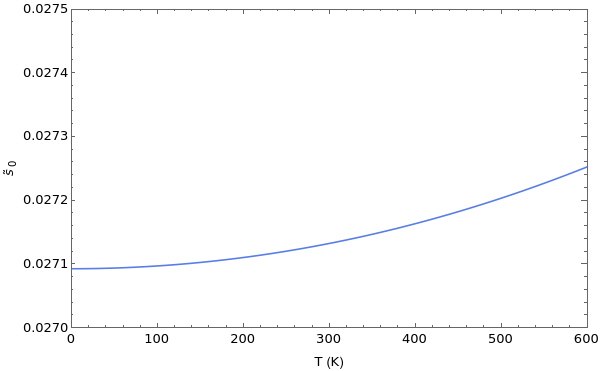}
  \end{subfigure}%
\begin{subfigure}{.48\textwidth}
  \centering
  \includegraphics[width=\linewidth]{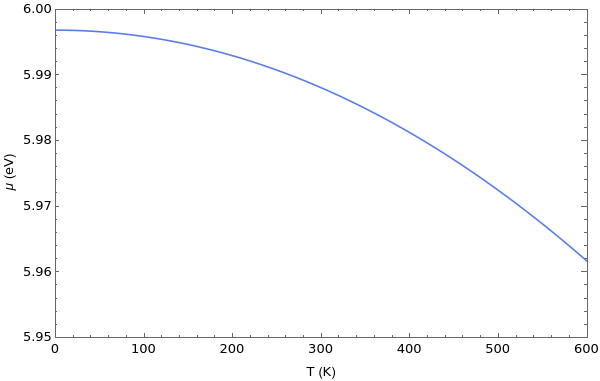}
\end{subfigure}
\caption{Value of the dimensionless parameters $\tilde s_0$ and the dimensionful chemical potential of the holographic theory $\mu$ chosen such that the plasma frequency of a layered system is fixed at all temperatures. We see that these value vary very little with temperature, with $\tilde s_0(T = 0) \approx 0.027$ and $\mu (T = 0) \approx 6 \text{eV}$}\label{fig:s0_mu_vs_T}
\end{figure}
\end{center}
\end{widetext}

\bibliography{apssamp}% Produces the bibliography via BibTeX.

\end{document}